\documentclass[%
pre, % two columns
aps,
amsmath,amssymb,floatfix,
reprint,%
twocolumn,%
longbibliography,
]{revtex4-1} 

\usepackage{amsmath,amsthm,latexsym,amssymb,amsfonts,epsfig}

\usepackage[english]{babel}
\usepackage[utf8]{inputenc}			
\usepackage{graphicx, texdraw}   %for figures
\usepackage{bm}

\usepackage{lmodern}
\usepackage{mathtools}
\usepackage{mathrsfs} 

\usepackage[left=1.29cm, right=1.29cm, top=2.cm, bottom=2.cm]{geometry}     % preset margins

\usepackage{enumerate}

\usepackage[colorlinks=true,citecolor=cyan, linkcolor=blue]{hyperref}

\usepackage{csquotes} 

\usepackage[usenames, dvipsnames]{color}

\usepackage{amsmath}
\usepackage{amsfonts}
\usepackage{amssymb}
\usepackage{braket}

\usepackage[normalem]{ulem} %to strike the words 
\newcommand{\vect}[1]{\boldsymbol{#1}}

\newcommand{\boldNabla}{\boldsymbol{\nabla}}

\newcommand{\boldSigma}{\boldsymbol{\sigma}}

\newcommand{\kS}{\kappa_\mathrm{S}}
\newcommand{\kA}{\kappa_\mathrm{A}}
\newcommand{\kB}{\kappa_\mathrm{B}}

\newcommand{\alphaB}{\alpha_\mathrm{B}}

\newcommand{\R}{\vect{r}}
\newcommand{\Intd}{\mathrm{d }}

\newcommand{\F}{\vect{F}}

\newcommand{\X}{\vect{x}}

\newcommand{\Faxen}{Fax\'{e}n}

\newcommand{\eR}{\vect{e}_r}

\newcommand{\Gmatr}{\boldsymbol{\mathcal{G}}}

\newcommand{\vStok}{\vect{v}^{\mathrm{S}}}

\newcommand{\xOne}{\vect{x}_1}
\newcommand{\xTwo}{\vect{x}_2}

\newcommand{\vecD}{\vect{d}}
\newcommand{\infSum}{\sum_{n=0}^{\infty}}
\newcommand{\infSumOne}{\sum_{n=1}^{\infty}}

\newcommand{\bgamma}{\boldsymbol{\gamma}}
\newcommand{\vecT}{\vect{t}}
\newcommand{\dd}{\vect{d}}

\definecolor{niceRed}{rgb}{0.8, 0.2, 0.1}

\usepackage{epstopdf}

\usepackage{type1ec} %

\usepackage{wasysym}

\begin{document}
\title{Creeping motion of a solid particle inside a spherical elastic cavity.\ {II.} Asymmetric motion}

\author{Christian Hoell}
\email{christian.hoell@uni-duesseldorf.de }
\affiliation
{Institut f\"{u}r Theoretische Physik II: Weiche Materie, Heinrich-Heine-Universit\"{a}t D\"{u}sseldorf, Universit\"{a}tsstra\ss e 1, 40225 D\"{u}sseldorf, Germany}

\author{Hartmut L\"{o}wen}
\affiliation
{Institut f\"{u}r Theoretische Physik II: Weiche Materie, Heinrich-Heine-Universit\"{a}t D\"{u}sseldorf, Universit\"{a}tsstra\ss e 1, 40225 D\"{u}sseldorf, Germany}

\author{Andreas M. Menzel}
\affiliation
{Institut f\"{u}r Theoretische Physik II: Weiche Materie, Heinrich-Heine-Universit\"{a}t D\"{u}sseldorf, Universit\"{a}tsstra\ss e 1, 40225 D\"{u}sseldorf, Germany}

\author{Abdallah Daddi-Moussa-Ider}
\email{abdallah.daddi.moussa.ider@uni-duesseldorf.de}
\affiliation
{Institut f\"{u}r Theoretische Physik II: Weiche Materie, Heinrich-Heine-Universit\"{a}t D\"{u}sseldorf, Universit\"{a}tsstra\ss e 1, 40225 D\"{u}sseldorf, Germany}

\date{\today}

\begin{abstract}

An analytical method is proposed for computing the low-Reynolds-number hydrodynamic mobility function of a small colloidal particle asymmetrically moving inside a large spherical elastic cavity, the membrane of which is endowed with resistance toward shear and bending.
In conjunction with the results obtained in the first part [Daddi-Moussa-Ider, L\"{o}wen, and Gekle, Eur.\ Phys.\ J.\ E \textbf{41}, 104 (2018)], in which the axisymmetric motion normal to the surface of an elastic cavity is investigated, the general motion for an arbitrary force direction can now be addressed.
The elastohydrodynamic problem is formulated and solved using the classic method of images through expressing the hydrodynamic flow fields as a multipole expansion involving higher-order derivatives of the free-space Green's function.
In the quasi-steady limit, we demonstrate that the particle self-mobility function of a particle moving tangent to the surface of the cavity is larger than that predicted inside a rigid stationary cavity of equal size.
This difference is justified by the fact that a stationary rigid cavity introduces additional hindrance to the translational motion of the encapsulated particle, resulting in a reduction of its hydrodynamic mobility.
Furthermore, the motion of the cavity is investigated, revealing that the translational pair (composite) mobility, which linearly couples the velocity of the elastic cavity to the force exerted on the solid particle, is solely determined by membrane shear properties. 
Our analytical predictions are favorably compared with fully-resolved computer simulations based on a completed-double-layer boundary integral method.
\end{abstract}

\maketitle

\section{Introduction}

Many industrial and biological transport processes on the microscale predominantly occur under confinement, where hydrodynamic interactions with boundaries drastically alter the diffusive behavior of microparticles in viscous media.
Prime examples include particle sorting in microfabricated fluidic devices~\cite{stone04, fu99, lu04, huh05, schmid14}, membrane separation and purification in pharmaceutical industry~\cite{darvishmanesh11, adamo13, gutmann15}, as well as intracellular drug delivery and targeting via multifunctional nanocarriers, which release therapeutic agents in disease regions such as tumor or inflammation sites~\cite{colson12, hillaireau09, liu16, maeda13, naahidi13, rosenholm10, singh09, bareford07}. 
The uptake by cell membranes occurs via endocytosis or by direct penetration to reach target cellular compartments.

At small length scales, fluid flows are characterized by small Reynolds numbers, implying that viscous forces dominate inertial forces.
In these situations, the fluid-mediated hydrodynamic interactions are fully encoded in the mobility tensor, which linearly couples the velocities of microparticles to the forces and torques exerted on them~\cite{happel12, kim13, leal80}.
Even for simple geometric confinements, finding closed analytical solutions of diverse flow problems can be challenging.
Most theoretical approaches are based on the method of images, consisting of a set of (typically higher-order) singularities that are required to satisfy the prescribed boundary conditions at the confining boundaries~\cite{blake71}.
Using this approach, the solution of the Stokes equations in the presence of a point force singularity acting in a fluid domain bounded by a rigid spherical cavity has been obtained by Oseen~\cite{oseen28}.
Extensions of Oseen's solution have further been proposed~\cite{butler53,collins54,hasimoto56, hasimoto92, hasimoto97,shail87, shail88,sellier08}.
A particularly more compliant solution that separately considers both axisymmetric and asymmetric Stokeslets has later been presented by Maul and Kim~\cite{maul94, maul96}.
Meanwhile, the hydrodynamic coupling and rotational mobilities have been calculated for point-like particles~\cite{felderhof12b}.
In this context, the low-Reynolds-number swimming inside spherical containers has also attracted some attention~\cite{tsemakh04, lavrenteva05, reigh17, zhu17, reigh17prfluids, shaik18}.

In this manuscript, we examine the slow translational motion of a small colloidal particle moving inside a large spherical elastic cavity (that itself is floating in an infinitely-extended viscous fluid).
This setup may be viewed as a relevant model system for transport processes within biological media, such as elastic cell membranes. 
The cavity membrane is modeled as a two-dimensional hyperelastic sheet, endowed with resistance toward shear elasticity and bending rigidity.
This model has previously been employed to address the effect of elastic confinements on the diffusive behavior of colloidal particles moving close to planar~\cite{daddi16, daddi18epje} or curved elastic membranes~\cite{daddi17b,daddi17c, daddi17d, daddi17e}.

The present article is a natural extension of a preceding paper~\cite{daddi18cavity} (hereinafter referred to as part~I), where  the axisymmetric motion was examined.
The goal of the current study is to supplement and complement our previous results by quantifying the effect of the confining elastic cavity on the asymmetric motion of an encapsulated particle located at arbitrary position within the cavity.
Our approach is based on the method of images employed by Fuentes and collaborators~\cite{fuentes88,fuentes89}, who examined theoretically the hydrodynamic interactions between two unequally-sized spherical viscous drops at moderately small separations.
Our analytical investigations proceed through the calculation of the Green's functions associated with a point force acting inside a spherical elastic cavity.
The problem treated here does not possess the symmetry properties of the simpler axisymmetric case considered in part~I.
This makes it necessary to employ an alternative mathematical framework to obtain the solution of the flow problem for the asymmetric case. 
The calculated hydrodynamic flow field is 
used to determine the frequency-dependent mobility functions for an enclosed point particle.
This approximation is reasonable if the separation distance between the particle and the cavity surface is large compared to the particle size. 
Particularly, inside our deformable cavity, the mobility in the quasi-steady limit of vanishing frequency is shown to be always larger than the one predicted inside a rigid cavity with no-slip boundary condition.
Our theoretical results favorably compare to numerical simulations.

The remainder of the paper is organized as follows.
In sect.~\ref{sec:singularitySolution}, we use the multipole expansion method to find solutions of the elastohydrodynamic problem for the fluid inside and outside the cavity. 
We then provide in sect.~\ref{sec:mobility} analytical expressions of the hydrodynamic self-mobility function for a particle moving tangent to the surface of the cavity.
In sect.~\ref{sec:cavityMotionUndMemDeformation}, we assess the motion of the large cavity and determine the deformation field induced by the motion of the particle.
We provide in sect.~\ref{sec:conclusions} concluding remarks summarizing our findings.
The appendix contains explicit expressions for the series coefficients arising from the multipole expansion.

\section{Singularity solution}
\label{sec:singularitySolution}

\begin{figure}
	\includegraphics[scale=0.9]{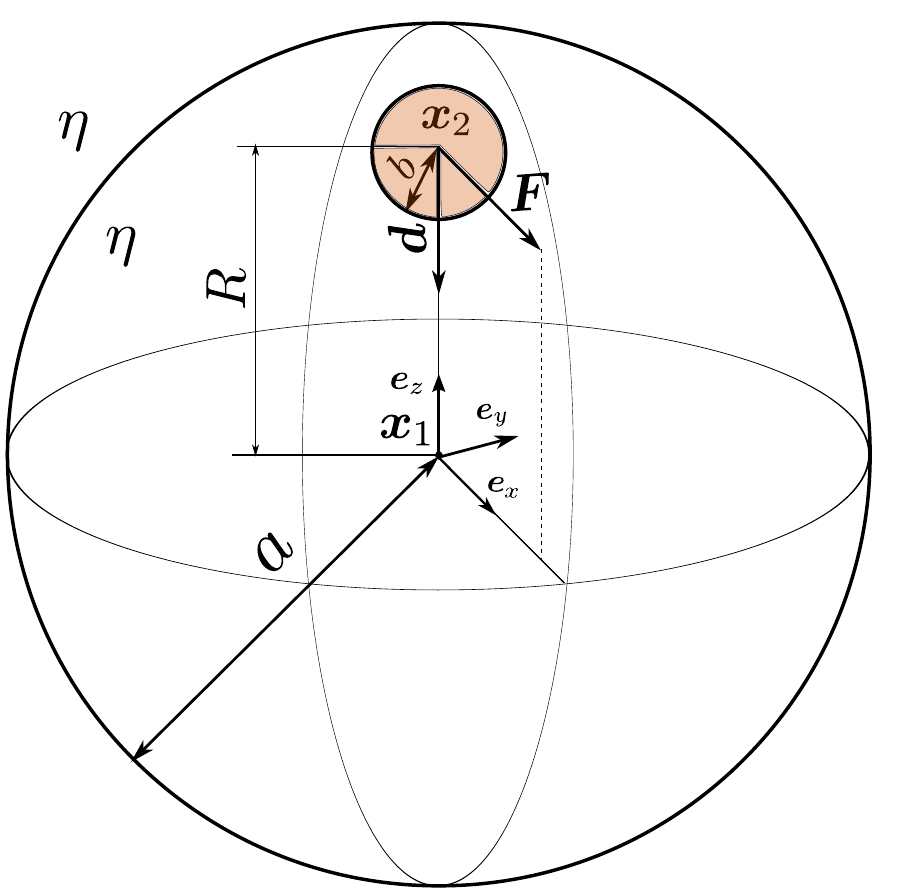}
	\caption{Graphical illustration of the system setup. A small spherical particle of radius~$b$ is located at~$\vect{x}_2 = R \vect{e}_z$ inside an elastic spherical cavity of radius~$a$ positioned at~$\vect{x}_1$.
	The fluid on both sides of the cavity is characterized by a constant dynamic viscosity~$\eta$.
	In an asymmetric configuration, the force is directed perpendicular to the unit vector~$\vecD = \left( \vect{x}_1 - \vect{x}_2 \right)/R$.}
	\label{illustration}
\end{figure}

We examine the low-Reynolds-number motion of a small sphere of radius~$b$ situated inside a large spherical elastic cavity of radius~$a$.
The fluid inside and outside the cavity is characterized by a constant dynamic viscosity~$\eta$, and the flow is assumed to be incompressible.
The center of the cavity at~$\vect{x}_1$ coincides with the origin of the spherical coordinate system.
The solid particle located at position~$\vect{x}_2 = R \vect{e}_z$ is moving under the action of an asymmetric external force~$\vect{F} \perp \vect{e}_z$.
An illustration of the system under consideration is shown in fig.~\ref{illustration}.

The physical problem is thus equivalent to solving the forced Stokes equations inside the cavity~\cite{happel12,kim13},
\begin{subequations}\label{stokesInside}
	\begin{align}
		\eta \boldNabla^2 \vect{v}^{(i)} - \boldNabla p^{(i)} + \vect{F} \delta \left( \vect{x}-\vect{x}_2 \right) &= 0 \, , \\
		\boldNabla \cdot \vect{v}^{(i)} &= 0 \, , 
	\end{align}
\end{subequations}
and homogeneous (force-free) equations for the outer fluid,
\begin{subequations}\label{stokesOutside}
	\begin{align}
		\eta \boldNabla^2 \vect{v}^{(o)} - \boldNabla p^{(o)} &= 0 \, , \\
		\boldNabla \cdot \vect{v}^{(o)} &= 0 \, , 
	\end{align}
\end{subequations}
wherein~$\vect{v}^{(i)}$ and~$\vect{v}^{(o)}$ denote the flow velocity fields for the inner and outer fluids, respectively, and $p^{(i)}$ and $p^{(o)}$ are the corresponding pressure fields.
Equations~\eqref{stokesInside} and~\eqref{stokesOutside} are subject to the regularity conditions 
\begin{subequations}
	\begin{align}
		\left| \vect{v}^{(i)} \right|<\infty &\text{~~as~~} r\to 0 \, , \\
		\vect{v}^{(o)} \to \vect{0} &\text{~~as~~} r\to \infty \, ,
	\end{align}
\end{subequations}
in addition to the standard boundary conditions of continuity of the velocity field and discontinuity of the hydrodynamic stresses at the cavity surface.
In the present work, we assume that the cavity undergoes a small deformation only, so that the boundary conditions are evaluated at the undeformed surface of reference at~$r=a$.
Specifically,
\begin{subequations}
	\begin{align}
		\left. \vect{v}^{(o)} - \vect{v}^{(i)} \right|_{r=a} &= \vect{0} \, , \label{Kontinuitaet} \\
		\left. \left( \boldSigma^{(o)} - \boldSigma^{(i)} \right) \cdot \eR \right|_{r=a} &= \Delta \vect{f}^\mathrm{S} + \Delta \vect{f}^\mathrm{B} \, , \label{Diskontinuitaet}
	\end{align}
\end{subequations}
where $\boldSigma = -p \vect{I} + 2\eta \vect{E}$ is the viscous stress tensor. Here, $\vect{E} =  \left( \boldNabla \vect{v} + \boldNabla \vect{v}^\mathrm{T} \right) / 2$ denotes the rate-of-strain tensor, the components of which are given in spherical coordinates by
\begin{subequations}
 \begin{align}
   \sigma_{\theta r} &= \eta \left( v_{\theta,r} - \frac{v_\theta + v_{r,\theta}}{r}  \right) \, , \label{Komponent_sigma_r_phi} \\
   \sigma_{\phi r} &= \eta \left( v_{\phi,r} + \frac{v_{r,\phi} - v_\phi}{r \sin\theta} \right) \, , \label{Komponent_sigma_r_theta} \\
   \sigma_{rr} &= -p + 2\eta v_{r,r} \, , \label{Komponent_sigma_r_r}
 \end{align}
\end{subequations}
where~$\phi$ and~$\theta$, respectively, denote the azimuthal and polar angles, such that $(\phi,\theta) \in [0,2\pi) \times [0,\pi]$ describes a point on the surface of the unit sphere.
Furthermore, by convention, indices after a comma stand for the corresponding partial derivatives, e.g., $v_{r,r}=\partial v_r / \partial r$.
Additionally, $\Delta \vect{f}^\mathrm{S}$ and~$\Delta \vect{f}^\mathrm{B}$ denote the traction jumps stemming from shear and bending deformation modes, respectively.
We further remark that, if the membrane cavity undergoes a large deformation, the boundary conditions should rather be evaluated at the displaced membrane positions, see, e.g., refs.~\onlinecite{sekimoto93, weekley06, salez15, saintyves16, rallabandi17, daddi18stone, rallabandi18}.

In this work, we model the elastic cavity as a spherical hyperelastic shell of vanishing thickness, the deformation of which is governed by the shear elasticity model proposed by Skalak~\cite{skalak73}
that is commonly employed when modeling, e.g., the membranes of red blood cells~\cite{Freund_2014, krueger12}.
Specifically, the areal strain energy density of the Skalak model is given by~\cite{krueger11}
\begin{equation}
	E = \frac{\kS}{12} \left(  (I_1^2 + 2I_1 - 2I_2) +C I_2^2 \right) \, ,
\label{W_SK}
\end{equation}
where $I_1$ and $I_2$ stand for the invariants of the right Cauchy-Green deformation tensor~\cite{green60, zhu14}, and $C=\kA/\kS$ is the Skalak coefficient representing the ratio between the area dilatation modulus~$\kA$ and shear modulus~$\kS$~\cite{skalak73}.
For $C = 1$, the Skalak model is equivalent to the classical Neo-Hookean model for small membrane deformations~\cite{lac04}.

Accordingly, the linearized traction jump due to shear is expressed in terms of the deformation field~$\vect{u}$, and can be split into an axisymmetric and an asymmetric part as
\begin{equation}
	\Delta \vect{f}^{\mathrm{S}} = \left. \Delta \vect{f}^{\mathrm{S}} \right|_{\mathrm{Axi}}  
	+ \left. \Delta \vect{f}^{\mathrm{S}} \right|_{\mathrm{Asy}}  
\end{equation}
where 
\begin{align}
		\left. \Delta f_\theta^\mathrm{S} \right|_\mathrm{Axi} &= 
	     -\frac{2\kS}{3} \bigg( 2\xi_{-} u_{r,\theta} 
	     + {\lambda} \left( u_{\theta,\theta\theta}  +  u_{\theta,\theta} \cot\theta \right)
	            \notag \\
	            &\quad- u_\theta \left( {\lambda} \cot^2 \theta + {\lambda}-1 \right) \bigg) \, , \notag \\
	   \left. \Delta f_\phi^\mathrm{S} \right|_\mathrm{Axi} &= 0 \, , \notag  \\ 
	    \left. \Delta f_r^\mathrm{S} \right|_\mathrm{Axi} &= 
	    \frac{4\kS}{3} \, \xi_- \left( 2u_r + u_{\theta,\theta} + u_{\theta} \cot \theta \right)  \, ,  \notag
\end{align}
and
\begin{align}
	 \left. \Delta f_\theta^\mathrm{S} \right|_\mathrm{Asy} &= 
	     -\frac{2\kS}{3} \bigg( \xi_{-}  \frac{u_{\phi,\phi\theta}}{\sin\theta} 
	            +\frac{u_{\theta,\phi\phi}}{2\sin^2\theta} 
	             - \xi_{+} \frac{\cot\theta}{\sin\theta} \, u_{\phi,\phi}  \bigg) \, , \notag \\
	  \left. \Delta f_\phi^\mathrm{S} \right|_\mathrm{Asy} &= -\frac{2\kS}{3} \bigg(
	  		\lambda \, \frac{ u_{\phi,\phi\phi}}{\sin^2\theta}
	  		+\frac{u_{\phi,\theta\theta}}{2} 
		    +\frac{\xi_-}{\sin\theta}
		    \left( 2u_{r,\phi} + u_{\theta,\phi\theta} \right)
	 	    \notag \\
	 	    &\quad+ \left(1-\cot^2\theta \right)
	 	     \frac{u_\phi }{2}
		    + \frac{u_{\phi,\theta}}{2} \cot\theta + \xi_+ \frac{\cot\theta}{\sin\theta} \, u_{\theta,\phi}   \bigg)   \, , \notag  \\ 
	  \left.  \Delta f_r^\mathrm{S} \right|_\mathrm{Asy} &= 
	    \frac{4\kS}{3} \frac{\xi_- }{\sin\theta} \, u_{\phi,\phi} \, ,  \notag
\end{align}
for the axisymmetric and asymmetric parts, respectively.
Here, the asymmetric part includes all terms that depend on $u_\phi$ or involve derivatives with respect to $\phi$.
Moreover, we have defined
\begin{subequations}
\begin{align}
  \lambda:&=1+C=1+\frac{\kA}{\kS} \, , \label{lambda} \\ 
  \xi_\pm &= \lambda \pm \frac{1}{2} \, .
\end{align}
\end{subequations}

In addition, we introduce a resistance toward bending following the Helfrich model~\cite{helfrich73,berndl90, seifert97}.
The areal bending energy density thus is described by a curvature-elastic continuum model of a quadratic form given by~\cite{GuckenbergerJPCM}
\begin{equation}
	E_\mathrm{B} =  2\kB \left( H-H_0 \right)^2,  \label{HelfrichHamiltonian}
\end{equation} 
wherein~$\kB$ denotes the bending modulus,
$H_0$ stands for the spontaneous curvature (here taken as the corresponding value for the initial undeformed sphere),
and $H:=b_\alpha^\alpha / 2$ (summing over repeated indices) is the mean curvature, with $b_\alpha^\beta$ being the corresponding component of the curvature tensor~\cite{kobayashi63}.

The traction jump equation across the membrane as derived from this model reads~\cite{GuckenbergerJPCM}
\begin{equation}
\Delta \vect{f} = -2\kB \left( 2(H^2-K+H_0 H) + \Delta_\parallel \right) (H-H_0) \, \vect{n} \, ,  \label{helfrich_tractionJump}
\end{equation}
where $\vect{n}$ is the outward-pointing unit normal vector to the spherical cavity,
$K:=\mathrm{det~} b_\alpha^\beta$ stands for the Gaussian curvature, and $\Delta_{\parallel}$ denotes the Laplace-Beltrami operator~\cite{deserno15}.
Accordingly, bending introduces a traction jump along the normal direction which can be split into an axisymmetric and an asymmetric part as
\begin{equation}
	\Delta f_{r}^{\mathrm{B}} = \left. \Delta f_{r}^{\mathrm{B}} \right|_{\mathrm{Axi}}  
	+ \left. \Delta f_{r}^{\mathrm{B}} \right|_{\mathrm{Asy}},
\end{equation}
where 
\begin{align}
	\left. \Delta f_{r}^{\mathrm{B}} \right|_{\mathrm{Axi}}  &= 
	\kB \bigg( 4u_r+ T\left(5+T^2\right)  u_{r,\theta} \notag \\
	&\quad+ \left(2-T^2\right) u_{r,\theta\theta} + 2T u_{r,\theta\theta\theta}  + u_{r, \theta\theta\theta\theta} \bigg) \, , \notag \\
	\left. \Delta f_{r}^{\mathrm{B}} \right|_{\mathrm{Asy}} &=
	\kB \left( 1+T^2 \right)
		              \Big( 2u_{r,\phi\phi\theta\theta} 
		              + 2\left(3+2T^2\right)u_{r,\phi\phi} \notag \\
		              &\quad- 2T u_{r,\phi\phi\theta} + \left(1+T^2\right)u_{r,\phi\phi\phi\phi} \Big) \, , \notag
\end{align}
and where we have used the shorthand notation~$T:=\cot\theta$.
We note that bending as derived from Helfrich's model does not introduce discontinuities along the tangential directions.
Accordingly, $\Delta f_{\theta}^{\mathrm{B}} = \Delta f_{\phi}^{\mathrm{B}} = \vect{0}$.
These traction jumps reduce to the axisymmetric case considered in part~I~\cite{daddi18cavity} for which $\left. \Delta \vect{f}^\mathrm{S} \right|_\mathrm{Asy} = \left. \Delta \vect{f}^\mathrm{B} \right|_\mathrm{Asy} = \vect{0}$.
In this situation, $u_\phi=0$ and all derivatives with respect to~$\phi$ drop out.

A closure of the problem is achieved by requiring a no-slip boundary condition at the undisplaced membrane.
Accordingly, the velocity field at $r=a$ is assumed to be equal to that of the displaced material points of the elastic cavity,
i.e., 
\begin{equation}
	\left. \vect{v} \right|_{r=a} = \frac{\Intd \vect{u}}{\Intd t} \, , 
\end{equation}
which can be written in Fourier space as
\begin{equation}
	\left. \vect{v} \right|_{r=a} = i\omega \, \vect{u} \, . \label{no-slip_Frequency}
\end{equation}

Our resolution methodology proceeds through writing the solution of the elastohydrodynamic problem inside the cavity as
\begin{equation}
	\vect{v}^{(i)} = \vect{v}^\mathrm{S} + \vect{v}^* \, , 
\end{equation}
where~$\vect{v}^\mathrm{S} = \boldsymbol{\mathcal{G}} \left(\X-\X_2\right) \cdot \vect{F}$ represents the velocity field induced by a point-force singularity acting at position~$\X_2$ in an unbounded fluid --- i.e., in the absence of the cavity --- and~$\vect{v}^*$ is the complementary term that is required to satisfy the imposed boundary conditions at the cavity. 
This type of complementary solution is often termed as the image system solution or sometimes known under the name of reflected flow field~\cite{blake71, blake74}.

We now briefly outline the main steps in our resolution approach.
First, we express
the Stokeslet solution in terms of harmonics, which are then rewritten in terms of harmonics relative to the origin via the Legendre expansion~\cite{zill11}.
Second, the reflected flow field and the solution outside the cavity are expressed 
using Lamb's general solution~\cite{lamb32} with interior and exterior harmonics, respectively.
This gives us a complete solution form involving a set of unknown series coefficients.
These coefficients are determined from the underlying boundary conditions imposed at the cavity surface.
Finally, the solution of the flow problem can then be employed to assess the effect of the confining cavity on the motion of the encapsulated spherical particle.

\subsection{Stokeslet representation}
\label{subsec:stokeslet}

For the remainder of this paper, we will scale all the lengths by the cavity radius~$a$.
In analogy with part~I, we begin by writing the Stokeslet singularity located at position~$\xTwo$ as
\begin{equation}
 \vStok = \Gmatr \left(\X-\X_2\right) \cdot\F = \frac{1}{8\pi\eta} \left(\frac{\boldsymbol{1}}{s} +  \frac{\vect{s} \vect{s}}{s^3} \right)\cdot\F \, , \label{stokeslet_at_X2}
\end{equation}
where we have defined $\vect{s} := \vect{x}-\xTwo$ and $s:= |\vect{s}|$.
Here, $\boldsymbol{1}$~is the unit tensor.
Using Legendre expansion, the harmonics located at $\X_2$ can conveniently be expressed in terms of harmonics centered at $\X_1$ as
\begin{equation}
 \frac{1}{s} = \infSum R^{n} \varphi_n(r,\theta) \, . \label{centered_harmonics}
\end{equation}
Here,~$\varphi_n$ are harmonics of degree~$n$, which are related to Legendre polynomials by~\cite{abramowitz72}
\begin{equation}
 \varphi_n (r, \theta) := \frac{(\dd \cdot \boldNabla)^n}{n!} \frac{1}{r} = \frac{1}{r^{n+1}} \, P_n (\cos \theta) \, , \notag
\end{equation}
where $\vecD := ( \xOne - \xTwo)/R$ is a unit vector, $\R= \X-\X_1$ is the position vector  in the spherical coordinate system centered at the cavity center, and $r := |\R|$. 
The dyadic product in eq.~\eqref{stokeslet_at_X2} can be written as
\begin{equation}
 \frac{\vect{s} \vect{s}}{s^3} = \vect{s} \, \boldNabla_2 \left(
\frac{1}{s}\right),
\end{equation}
with $\boldNabla_2 := {\partial }/{\partial \vect{x}_2 }$. 
By making use of eq.~\eqref{centered_harmonics}, the  derivatives with respect to $\X_2$ can readily be taken care of by noting that
\begin{equation}
 \boldNabla_2 R^n = -n R^{n-1} \vecD \, , \qquad \left(\vecD \cdot \boldNabla_2 \right) \vecD = \vect{0} \, .
\end{equation}

In the present work, we focus our attention on the asymmetric situation in which the force is purely tangent to the membrane surface and thus~$\vect{F}\cdot\vect{d}=0$. 
By taking this into consideration, the Stokeslet stated in eq.~\eqref{stokeslet_at_X2} can therefore be expressed as 
\begin{equation}
 \begin{split}
  8\pi\eta \vStok = &{} \F \infSum R^{n} \, \varphi_n 
  -\R \infSumOne R^{n-1} \left(\F\cdot\boldNabla\right)\varphi_{n-1} \\
  &-\dd \infSumOne R^{n} \left(\F\cdot\boldNabla\right)\varphi_{n-1} \, .
 \end{split} \notag
\end{equation}
Accordingly, the Stokeslet solution has now been expressed in terms of spherical harmonics positioned at the origin.
By defining $\vecT = \F \times \vecD$, we have the recurrence relation
\begin{equation}
 \vecD (\F\cdot\boldNabla) \varphi_n = (\vecT \times \boldNabla) \varphi_n + (n+1) \F  \varphi_{n+1} \, . \label{identityRotlet}
\end{equation}
In addition, imposing $\F\cdot\dd=0$ yields
\begin{equation}
	\begin{split}
		\label{eliminateFPhi}
		(2n+1)(n+1)\F\varphi_{n} &=
		-(2n+3)\R\psi_n  - {r^2} \boldNabla\psi_n \\
		&\quad\quad+\boldNabla\psi_{n-2} - (2n+1)\bgamma_{n-1}  \, ,  
	\end{split}
\end{equation}

where the harmonics~$\psi_n$ and~$\bgamma_n$ are, respectively, defined as
\begin{equation}
	\psi_n = (\F\cdot\boldNabla)\varphi_{n} \, , \qquad \bgamma_n = (\vecT\times\boldNabla)\varphi_{n} \, .
\end{equation}
These are related to each other via $\psi_n = \bgamma_n \cdot \vecD$.

In the following, the functions $\boldNabla\psi_n$, $\vect{r}\psi_n$, and $\bgamma_n$ are chosen as vector basis functions to be used for expanding the velocity and pressure fields.
Accordingly, the Stokeslet solution can be written in a final form as
\begin{align}
 8\pi\eta\vStok &= \infSumOne \left( \frac{\left(n-2\right) R^{n-1}}{(2n-1){n}} \, r^2 
                 - \frac{n R^{n+1}}{(n+2)(2n+3)} \right) \boldNabla\psi_{n-1} \notag \\ 
                 &-\frac{2 R^n}{n+1} \, \bgamma_{n-1} 
                 - \frac{2(n+1) R^{n-1}}{n(2n-1)} \, \R\psi_{n-1} \, . \label{Stokeslet_finalize}
\end{align}

We next proceed to deriving analogous expansions for the flow fields inside and outside the spherical cavity.

\subsection{The image system solution}
\label{subsec:image}

The solution for the flow field in a spherical domain possesses a generic form known as Lamb's general solution~\cite{lamb32,kim13}. 
It involves three sets of unknown coefficients to be determined from the underlying boundary conditions, and can be written for an asymmetric situation as
\begin{equation}
		8\pi\eta\vect{v}^{*} =  \infSumOne 
		\left( a_n {\vect{\sigma}_n}_1 + b_n {\vect{\sigma}_n}_2 + c_n {\vect{\sigma}_n}_3 \right) \, , \label{insideSolution_finalized}
\end{equation}
where we have defined 
	\begin{align}
		{\vect{\sigma}_n}_1 &= \frac{n+3}{2n} \, r^{2n+3} \boldNabla\psi_{n-1}  
				+ \frac{(n+1)(2n+3)}{2n} \, r^{2n+1} \R \psi_{n-1} \, , \notag \\
		{\vect{\sigma}_n}_2 &= \frac{r^{2n+1}}{n}\boldNabla\psi_{n-1} 
				+ \frac{2n+1}{n} \, r^{2n-1}\R \psi_{n-1} \, , \notag \\
		{\vect{\sigma}_n}_3 &= r^{2n-1} \bgamma_{n-1} 
				+ (2n-1) r^{2n-3} (\vecT\times\R) \varphi_{n-1} \, . \notag	
	\end{align}

Here, $a_n$, $b_n$, and~$c_n$ are free parameters that will be determined from the boundary conditions.
It is worth noting that the present solution involves three unknown coefficients for each $n$,
while the simpler axisymmetric motion considered in part~I only involves two sets of coefficients.
Unfortunately, this also means that we are not able to to proceed as for the axisymmetric case, but have to derive the solutions using a notably different framework.

\subsection{The exterior solution}
\label{subsec:exterior}

The solution on the outside of the spherical cavity can be expressed in terms of exterior harmonics using Lamb's general solution as
\begin{equation}
	\begin{split}
		8\pi\eta \vect{v}^{(o)} &= \infSumOne  
			\Bigg( A_n \left( \frac{n-2}{2(n+1)} \, r^2 \boldNabla \psi_{n-1} - \vect{r} \psi_{n-1}  \right) \\
			&-\frac{B_n}{n+1} \, \boldNabla \psi_{n-1}
			+ C_n \bgamma_{n-1} \Bigg) \, . \label{outsideSolution_finalized}
	\end{split}
\end{equation}
The latter expression can be deduced from the solution for the inner fluid given by eq.~\eqref{insideSolution_finalized} by making use of the substitution $n \leftarrow -(n+1)$.

The six unknown coefficients ($a_n$, $b_n$, and~$c_n$ for the image system solution, and $A_n$, $B_n$, and~$C_n$ for the exterior flow) can now be determined from the underlying boundary conditions of continuity of the flow velocity field and discontinuity of the hydrodynamic stress tensor across the membrane.

\subsection{Velocity projections}
\label{subsec:projections}

Before proceeding with the determination of the unknown series coefficients, it is convenient to state explicitly the projected expressions of the velocity field along the radial and tangential directions.

\subsubsection{Radial velocities}

The radial projection of the three vector basis functions are given by 
\begin{subequations}\label{radialProjectionRelations}
\begin{align}
  \eR\cdot\boldNabla\psi_{n-1} &= -\frac{n+1}{r} \, \psi_{n-1}\, , \label{radialProjectionRelation_1} \\ 
  \eR\cdot\R\psi_{n-1} &= r \psi_{n-1} \, , \\
  \eR\cdot\bgamma_{n-1} &= -\frac{1}{r} \, \psi_{n-2} \, .
\end{align}
\end{subequations}
In addition to that, since~$\eR$ and $\R$ are collinear, the scalar triple product $\eR\cdot(\vecT\times\R)\, \varphi_{n-1}$ vanishes.
Moreover, the projection of eq.~\eqref{eliminateFPhi} onto the radial direction yields
\begin{equation}
 \eR \cdot \F \varphi_{n} = \frac{1}{2n+1} \left( \frac{\psi_{n-2}}{r} - r \psi_n \right) \, . \label{latestRadialProjectionRelation}
\end{equation}

By making use of eqs.~\eqref{radialProjectionRelations} and~\eqref{latestRadialProjectionRelation} in the radial projection of eqs.~\eqref{Stokeslet_finalize}, \eqref{outsideSolution_finalized}, and \eqref{insideSolution_finalized}, the components of the fluid velocity fields along the radial direction can thus be expressed in terms of the harmonics $\psi_n$ as
\begin{subequations}\label{radialVelo}
	\begin{align}
		8\pi\eta v_r^\mathrm{S} &= \infSumOne
		\left( \frac{n+3}{2n+3} \frac{R^2}{r^2} - \frac{n+1}{2n-1} \right) R^{n-1} r \psi_{n-1} \, , \\
		8\pi\eta v_r^{*} &= \infSumOne 
		\left( \frac{n+1}{2} \, a_n r^{2} 
		+ b_n - c_{n+1} \right) r^{2n} \psi_{n-1} \, , \\
		8\pi\eta v_r^{(o)} &= \infSumOne 
		\left( -\frac{nr}{2} \, A_n + \frac{B_n}{r} - \frac{C_{n+1}}{r} \right) \psi_{n-1} \, .
	\end{align}
\end{subequations}

\subsubsection{Tangential velocities}

As for the tangential direction, we define the projection operator $\boldsymbol{\Pi} := \boldsymbol{1} - \eR \eR $, which projects vectors on a plane tangent to the surface of the spherical cavity. 
By applying the projection operator to eq.~\eqref{eliminateFPhi}, we readily obtain
\begin{equation}
 (\boldsymbol{\Pi} \F) \varphi_n = \frac{1}{n+1} \bigg( \frac{1}{2n+1} \left( \boldsymbol{\Psi}_{n-2} - r^2 \boldsymbol{\Psi}_n \right)-\boldsymbol{\Gamma}_{n-1} \bigg) \, , \label{tangentialProjectionRelation_1}
\end{equation}
where we have defined the vector harmonics
\begin{equation}
 \boldsymbol{\Gamma}_n := \boldsymbol{\Pi} \bgamma_n \, , \quad\quad \boldsymbol{\Psi}_n := \boldsymbol{\Pi} \boldNabla \psi_n \, . \notag
\end{equation}
Additionally, the tangential projection of~$\left(\vecT \times \R\right)\varphi_n$ can be taken care of by noting that
\begin{equation}
 \begin{split}
  \label{tangentialProjectionRelation_2}
     \boldsymbol{\Pi} (\vecT\times\R)\varphi_{n-1} &= \frac{1}{2n-1} \bigg( \frac{1}{n-1} \left( \boldsymbol{\Psi}_{n-4} - r^2 \boldsymbol{\Psi}_{n-2} \right) \\
     &\quad-\frac{n-2}{n-1} \, \boldsymbol{\Gamma}_{n-3} - r^2 \, \boldsymbol{\Gamma}_{n-1} \bigg) \, . 
 \end{split}
\end{equation}
Applying the projection relations stated by eqs.~\eqref{tangentialProjectionRelation_1} and \eqref{tangentialProjectionRelation_2} to eqs.~\eqref{Stokeslet_finalize}, \eqref{outsideSolution_finalized}, and \eqref{insideSolution_finalized}, we finally obtain
\begin{widetext}
\vspace{-20pt}
\begin{subequations}\label{tangentialVelo}
	\begin{align}
	   8\pi\eta \, \boldsymbol{\Pi} \vStok &= \infSumOne \left( \frac{\left(n-2\right) R^{n-1}}{(2n-1){n}} \, r^2 - \frac{n R^{n+1}}{(n+2)(2n+3)} \right) \boldsymbol{\Psi}_{n-1}
	   + \infSum 	-\frac{2R^{n+1}}{n+2} \, \boldsymbol{\Gamma}_{n} \, , \\
	   8\pi\eta \, \boldsymbol{\Pi} \vect{v}^{*} &= \infSumOne \bigg( \frac{r^{2n+3}}{n+2} \, c_{n+3}-\frac{r^{2n+1}}{n} \, c_{n+1}+ \frac{r^{2n+1}}{n} \, b_n
	   + \frac{n+3}{2n} \, r^{2n+3} a_n \bigg) \boldsymbol{\Psi}_{n-1} + \infSum -\frac{n+1}{n+2} \, r^{2n+3} c_{n+3} \, \boldsymbol{\Gamma}_n \, ,  \label{vInside_t} \\
	   8\pi\eta \, \boldsymbol{\Pi} \vect{v}^{(o)} &= \infSumOne  
	   \frac{1}{n+1} \left( \frac{n-2}{2} \, r^2 A_n - B_n \right) \boldsymbol{\Psi}_{n-1}
	   + \infSum C_{n+1} \boldsymbol{\Gamma}_n \, .
	\end{align}
\end{subequations}
\vspace{-20pt}
\end{widetext}

% \newpage

\subsection{Determination of the series coefficients}
\label{subsec:series}

To determine the unknown coefficients, we have to make recourse to the orthogonality properties of spherical harmonics~\cite{edmonds96}. 
For this purpose, we introduce the following notation to describe the average of a given quantity~$Q(\phi,\theta)$ over the surface of a sphere.
Specifically, this means
\begin{equation}
\langle Q \rangle := \frac{1}{2\pi} \int_0^{2\pi} \int_0^\pi Q (\phi,\theta) \sin \theta \, \Intd \theta \, \Intd \phi \, . \label{averagingDefinition}
\end{equation}

At the surface of the cavity, i.e., for $r=1$, the harmonics~$\varphi_n$ and~$\psi_n$ satisfy the orthogonality relations
\begin{subequations}
	\begin{align}
	 \left. \langle \varphi_{m-1} \varphi_{n-1} \rangle \right|_{r=1} &= \frac{2}{2n+1} \, \delta_{mn} \, , \\
	 \left. \langle \psi_{m-1} \psi_{n-1} \rangle \right|_{r=1} &= \frac{n(n+1)}{2n+1} \, \delta_{mn} \, , 
	\end{align}
\end{subequations}
where~$\delta_{mn}$ denotes the Kronecker symbol, i.e., the above terms vanish for $m \neq n$.
Moreover, the vector harmonics $\boldsymbol{\Psi}_{n-1}$ and $\boldsymbol{\Gamma}_n$ satisfy at $r=1$ the orthogonality properties
\begin{subequations} \label{MainOrthoEqs}
	\begin{align}
	 \left. \langle \boldsymbol{\Psi}_{m-1} \cdot \boldsymbol{\Psi}_{n-1}  \rangle \right|_{r=1} &= \frac{n^2(n+1)^2}{2n+1} \, \delta_{mn} \, , \label{Psi_DOT_Psi} \\
	 \left. \langle \boldsymbol{\Gamma}_m \cdot \boldsymbol{\Gamma}_n  \rangle \right|_{r=1} &=\frac{4(n+1)^3}{(2n+1)(2n+3)} \, \delta_{mn} \, , \label{Gamma_DOT_Gamma} \\
	 \left. \langle \boldsymbol{\Psi}_{m-1} \cdot \boldsymbol{\Gamma}_n \rangle \right|_{r=1} &= \frac{n^2(n+1)}{2n+1} \, \delta_{mn} \, . \label{Psi_DOT_Gamma}
	\end{align}
\end{subequations}
We further note that their derivatives with respect to~$r$ (needed in the calculation of the stress jumps) satisfy the recurrence relations
\begin{subequations}
	\begin{align}
	 \left. \left( \boldsymbol{\Psi}_{n-1,r}  + (n+2) \boldsymbol{\Psi}_{n-1}\right) \right|_{r=1} &= 0 \, ,  \\
	 \left. \left( \boldsymbol{\Gamma}_{n,r}  + (n+2) \boldsymbol{\Gamma}_n  \right) \right|_{r=1} &= 0\, . 
	\end{align}
\end{subequations}

\subsubsection{Pressure field}

Knowing the velocity fields on both sides of the elastic cavity, the inner and outer pressure fields can readily be calculated from the fluid motion equations.
The solution inside the spherical cavity, which comprises both contributions from the Stokeslet and the image system solution, can be expressed in terms of a multipole expansion as
\begin{equation}
	8\pi p^{(i)} = \infSumOne \left( -2R^{n-1} + \frac{(n+1)(2n+3)}{n} \, r^{2n+1} a_n  \right) \psi_{n-1} \, . \notag
\end{equation}
Outside the cavity, only the exterior harmonics that decay at larger distances should be accounted for, thus excluding contributions of the form~$r^{2n+1} \psi_{n-1}$.
After some algebra, we obtain
\begin{equation}
	8\pi p^{(o)} = \infSumOne - \frac{n(2n-1)}{n+1} \, A_n \psi_{n-1} \, . \notag
\end{equation}

\subsubsection{Continuity of velocity}

The projections of the fluid velocity field along the radial and tangential directions can be presented in a generic form as
\begin{subequations}
	\begin{align}
		v_r^{(q)}                        &= \infSumOne \rho_n^{(q)} \psi_{n-1} \, , \\
		 \boldsymbol{\Pi} \vect{v}^{(q)} &= \infSumOne \alpha_n^{(q)} \boldsymbol{\Psi}_{n-1} + \infSum \beta_n^{(q)} \boldsymbol{\Gamma}_n \, ,  
	\end{align}
\label{fluid_flows}
\end{subequations}
wherein~$q=i$ holds for the fluid on the inside, and $q=o$ for the fluid on the outside.
Moreover, $\rho_n^{(q)}$, $\alpha_n^{(q)}$, and~$\beta_n^{(q)}$, for $q \in \{i,o\}$, are radially symmetric series functions that can readily be obtained by identification with eqs.~\eqref{radialVelo} and~\eqref{tangentialVelo} giving the radial and tangential velocities, respectively.

The unknown coefficients inside the cavity can conveniently be expressed in terms of those outside thanks to the natural continuity of the velocity field across the membrane. 
By making use of the orthogonality properties of the basis functions, we obtain
\allowdisplaybreaks
\begin{subequations}\label{CoeffsOnTheInside}
	\begin{align}
		a_n &= \frac{n(2n-1)}{2(n+1)} \, A_n - \frac{2n+1}{n+1} \, B_n + \frac{2n+1}{n+1} \, C_{n+1} \notag \\
		&+ R^{n-1} \left( \frac{(n+3)(2n+1)}{(n+1)(2n+3)} \, R^2 - 1 \right) \, ,  \\
		b_n &= -\frac{n(2n+1)}{4} \, A_n + \frac{2n+3}{2} \, B_n - \frac{2n+3}{2} \, C_{n+1}  \\
		&- \frac{n \, C_{n-1}}{n-1} + R^{n-1} \left( \frac{2n^3+n^2-10n+3}{2(2n-1)(n-1)}  - \frac{n+3}{2} \, R^2 \right) , \notag \\
		c_n &= -\frac{(n-1) C_{n-2} + 2R^{n-2}}{n-2} \, . 
	\end{align}
\end{subequations}

\subsubsection{Discontinuity of stress tensor}

Expressions for the unknown coefficients~$A_n$, $B_n$, and~$C_n$ associated with the outer fluid can be obtained from the traction jump equations across the membrane.
For the sake of clarity, and to make the calculations traceable, we will consider in the following the effects of shear and bending deformation modes separately.

\paragraph{Pure shear} 

The tangential traction jump equations due to shear can conveniently be cast in the form
\begin{equation}\label{tangentialTractionJump_Shear}
 \begin{split}
  \infSumOne \tilde{\alpha}_n & \boldsymbol{\Psi}_{n-1} 
  + \left.\infSum \tilde{\beta}_n \boldsymbol{\Gamma}_n \right|_{r=1} = \\
 & \left. \infSumOne \alpha_n^{(o)} \vect{F}_n+\infSum \beta_n^{(o)} \vect{G}_n+\infSumOne \rho_n^{(o)} \vect{f}_n \right|_{r=1} \, , 
 \end{split}
\end{equation}
where we have defined
\begin{align}
	\tilde{\alpha}_n &= \alpha_{n,r}^{(o)} - \alpha_{n,r}^{(i)}-(n+2) \left( \alpha_n^{(o)} - \alpha_n^{(i)} \right) \, ,  \notag \\
	\tilde{\beta}_n &= \beta_{n,r}^{(o)} - \beta_{n,r}^{(i)}-(n+2) \left( \beta_n^{(o)} - \beta_n^{(i)} \right) \, . \notag
\end{align}
Here, $\vect{F}_n$, $\vect{G}_n$, and~$\vect{f}_n$ are known series vectors, the expressions of which can be obtained by identification with eq.~\eqref{Diskontinuitaet} upon substitution of the tangential velocity field from eq.~\eqref{tangentialVelo}.
They satisfy the orthogonality relations
\begin{align}
 \left. \langle \vect{F}_n\cdot \boldsymbol{\Psi}_{m-1} \rangle \right|_{r=1} &= n(n+1) \big( n(n+1){\lambda}-1 \big) S_n \delta_{mn} \, , \notag \\
 \left. \langle \vect{G}_n\cdot \boldsymbol{\Psi}_{m-1} \rangle \right|_{r=1} &= n \big( n(n+1){\lambda}-1 \big) S_n \delta_{mn} \, , \notag \\
 \left. \langle \vect{f}_n\cdot \boldsymbol{\Psi}_{m-1} \rangle \right|_{r=1} &= -n(n+1) \left( 2{\lambda}-1 \right) S_n \delta_{mn} \, , \notag
\end{align}
with the basis vector harmonics~$\boldsymbol{\Psi}_{m-1}$, and
\begin{align}
	 \left. \langle \vect{F}_n\cdot \boldsymbol{\Gamma}_m \rangle \right|_{r=1} &= n \big( n(n+1){\lambda}-1 \big) S_n \delta_{mn} \, , \notag \\
	 \left. \langle \vect{G}_n\cdot \boldsymbol{\Gamma}_{m} \rangle \right|_{r=1} &= \frac{S_n W_n}{2n+3}   \, \delta_{mn} \, , \notag \\
	 \left. \langle \vect{f}_n\cdot \boldsymbol{\Gamma}_{m} \rangle \right|_{r=1} &= -n \left( 2{\lambda}-1 \right) S_n \delta_{mn} \, , \notag
\end{align}
with~$\boldsymbol{\Gamma}_n$, where we have defined
% \begin{subequations}
	\begin{align}
		S_n &=  \frac{\alpha n(n+1)}{2n+1} \, , \notag \\
		W_n &= 6+11n+\tfrac{13}{2}\,  n^2+n^3+n^2(2n+3)\lambda \, , \notag 
	\end{align}
% \end{subequations}
with
\begin{equation}
	\alpha = \frac{2\kS}{3\eta i\omega} 
\label{alpha}
\end{equation}
being the shear number.
Combining these equations with the orthogonality relations given by eqs.~\eqref{MainOrthoEqs} yields
\begin{widetext}
\begin{subequations}\label{wideEqsShear}
	\begin{align} 
		 \left. \tilde{\alpha}_n + \frac{\tilde{\beta}_n}{n+1} \right|_{r=1} &= 
		 \left. \alpha \Bigg( \left( \alpha_n^{(o)} + \frac{\beta_n^{(o)}}{n+1} \right) 
		 \big( n(n+1){\lambda}-1 \big) - (2{\lambda}-1) \rho_n^{(o)} \Bigg) \right|_{r=1} \, , \\
		 \left. \tilde{\alpha}_n +\frac{4(n+1)^2}{(2n+3)n^2} \, \tilde{\beta}_n \right|_{r=1} &= \left. \alpha \Bigg( \big( n(n+1){\lambda}-1 \big) \alpha_n^{(o)} - (2{\lambda}-1)\rho_n^{(o)}
		 + \left( \frac{12+22n+13n^2+2n^3}{2n(2n+3)} + \lambda n \right) \beta_n^{(o)}  \Bigg) \right|_{r=1} \, . 
		\end{align}
\end{subequations}

Using our representation, the normal traction jump due to shear reads
\begin{equation}
 	\begin{split}
 		\left. \infSumOne  \big( p_n^{(o)} - p_n^{(i)} \big) \psi_{n-1} \right|_{r=1} = 
 		\left. \alpha (2{\lambda}-1) \infSumOne \left( \rho_{n,r}^{(o)} - (n+1)\rho_n^{(o)} \right) \psi_{n-1} \right|_{r=1} \, , 
 	\end{split}
\end{equation}
\end{widetext}
which, upon using the orthogonality property of~$\psi_{n-1}$, yields
\begin{equation}
 \left. p_n^{(o)} - p_n^{(i)} \right|_{r=1} = \left. \alpha (2{\lambda}-1) \left( \rho_{n,r}^{(o)} - (n+1)\rho_n^{(o)} \right) \right|_{r=1} \, . \label{eq_6}
\end{equation}

By combining eqs.~\eqref{CoeffsOnTheInside}, \eqref{wideEqsShear}, and \eqref{eq_6}, the unknown series coefficients for the outer fluid can be obtained and cast in the form
\begin{subequations}\label{AnBnCnShear}
	\begin{align}
	  A_n &=-\frac{( n+1 ) ( 2n+1 )}{K_3} \left( K_1 \, R^{n+1} +  K_2 \, R^{n-1} \right) \, ,  \\
	  B_n &= \frac{K_4}{K_5} \, A_n + \frac{1}{K_7} \left( K_6 \, R^{n+1} +  K_8 \, R^{n-1} \right) \, ,  \\
	C_n &=- \frac {2n+1}{K_9} \, {R}^{n} \, , 
	 \end{align}
\end{subequations}
where~$K_1, \dots, K_9$ are rather complex functions of~$\alpha, \lambda$ and~$n$, the expressions of which are explicitly provided in the Appendix. 
In the limit~$i \alpha\to\infty$, which physically corresponds to a cavity membrane with an infinite shear elasticity modulus (or equivalently to a vanishing actuation frequency), the expressions of the series coefficients inside the cavity reduce to 
\begin{subequations}\label{anbncnRigid}
	\begin{align}
		\lim_{\alpha\to\infty} a_n &= \frac{(n+3)(2n+1)}{(n+1)(2n+3)} \, R^{n+1} - R^{n-1} \, , \\
		\lim_{\alpha\to\infty} b_n &= \frac{2n^3+n^2-10n+3}{2(n-1)(2n-1)} \, R^{n-1} 
										- \frac{n+3}{2} \, R^{n+1} \, , \\
		\lim_{\alpha\to\infty} c_n &= -\frac{2}{n-2} \, R^{n-2} \, .
	\end{align}
\end{subequations}	
In this limit, $A_n$, $B_n$, and~$C_n$ vanish except for $n=1$, where $(A_1,B_1,C_1) = (4,2/3,-R)$.
It is worthwhile to note that the coefficients given by eqs.~\eqref{anbncnRigid} correspond to the solution for an asymmetric point force acting inside a rigid cavity with no-slip boundary conditions.

\paragraph{Pure bending}

We now use a similar resolution procedure to determine the unknown series coefficients for a cavity membrane with pure bending resistance, such as that of a fluid vesicle or a liposome used as a vehicle for pharmaceutical drugs~\cite{rui98, torchilin05, zylberberg16}.
Since the tangential components of the traction are continuous, we obtain
\begin{equation}\label{tangentialTractionJump_Bending}
 \left. \infSumOne \tilde{\alpha}_n \boldsymbol{\Psi}_{n-1} +\infSum \tilde{\beta}_n \boldsymbol{\Gamma}_n \right|_{r=1} = \vect{0} \, , 
\end{equation}
which, after applying the orthogonality properties given by eqs.~\eqref{MainOrthoEqs}, leads to
\begin{align} 
 \tilde{\alpha}_n |_{r=1} = \tilde{\beta}_n |_{r=1} = 0 \, . \label{alphabeta0}
\end{align}

The normal traction jump due to bending as derived from the Helfrich model reads
\begin{equation}
	\infSumOne \left. \Big( p_n^{(o)} - p_n^{(i)} \Big) \psi_{n-1} \right|_{r=1} = 
	\left. \infSumOne -\rho_n^{(o)} H_n \right|_{r=1} \, ,
\end{equation}
which, upon using the orthogonality relation
\begin{equation}
	\langle H_n \psi_{m-1} \rangle |_{r=1} = \alphaB \frac{n(n+1)(n-1)^2 (n+2)^2}{2n+1} \, \delta_{mn} \, , 
\end{equation}
leads to
\begin{equation}
	\left. p_n^{(o)} - p_n^{(i)} \right|_{r=1} = \left. -\alphaB (n-1)^2 (n+2)^2 \rho_n^{(o)} \right|_{r=1} \, , \label{eq_6ben}
\end{equation}
wherein
\begin{equation}
	\alphaB = \frac{\kB}{\eta i\omega} 
\end{equation}
denotes the bending number.

By combining eqs.~\eqref{alphabeta0} and \eqref{eq_6ben} with eqs.~\eqref{CoeffsOnTheInside}, the unknown series coefficients for the fluid on the outside can be cast in the form
\begin{subequations}\label{AnBnCnBending}
	\begin{align}
		 A_n &= \frac{n+1}{Q_3} \left( Q_1  R^{n+1} + Q_2 R^{n-1} \right) \, ,\\
		 B_n &= \frac{1}{Q_7} \left( Q_4 A_n + Q_5 R^{n+1} + Q_6 R^{n-1} \right) \, ,\\
		 C_n &=-\frac{2}{n+1} \, R^n \, ,
	\end{align}
\end{subequations}
where $Q_1, \dots, Q_7$ are complicated functions of $\alpha_\mathrm{B}$, $\lambda$, and $n$ which are given in the Appendix.
In the limit~$i\alphaB\to\infty$, corresponding to an infinite membrane bending modulus, or to a vanishing forcing frequency, the series coefficients are given by
\begin{subequations}
	\begin{align}
		\lim_{\alphaB\to\infty} a_n &= 
		\frac{(n+3)(2n-1)}{2(n+1)(2n+3)} \, R^{n+1} - \frac{R^{n-1}}{2} \, , \\
		\lim_{\alphaB\to\infty} b_n &=
		-\frac{n+3}{4} \, R^{n+1} + \frac{(n+1)(2n+3)}{4(2n-1)} \, R^{n-1} \, , \\ 
		\lim_{\alphaB\to\infty} c_n &= 0 %\, , 
	\end{align}
\end{subequations}
for the inner fluid, and
\begin{subequations}
	\begin{align}
			\lim_{\alphaB\to\infty} A_n &= 
			\frac{1}{2n} \left( (n+3)R^{n+1} - (n+1)R^{n-1} \right) \, , \\
			\lim_{\alphaB\to\infty} B_n &=
			-\frac{n+1}{4} \, R^{n-1} + \frac{n^2+5n-2}{4(n+2)} \, R^{n+1} \, , \\
			\lim_{\alphaB\to\infty} C_n &=
			-\frac{2}{n+1} \, R^n %\, ,
	\end{align}
\end{subequations}
for the outer fluid when~$n \ge 2$.
In addition, $(A_1,B_1,C_1) = (4, 2R^2/15, -R)$.

\paragraph{Combined shear and bending}

An analogous resolution strategy can be adopted for the determination of the sum coefficients when the membrane is simultaneously endowed with both a resistance toward shear and bending.
Analytical expressions of the coefficients can readily be obtained using computer algebra systems but these are not provided here due to their complexity and lengthiness.
It is noteworthy that, in contrast to planar elastic membranes, a coupling between shear and bending deformation modes has been observed for curved membranes.

\section{Hydrodynamic mobility}
\label{sec:mobility}

The calculation of the flow field presented in the previous section can be utilized to assess the effect of the confining cavity on the motion of the encapsulated particle.
This effect is quantified by the hydrodynamic self-mobility function~$\mu$, which relates the translational velocity of a colloidal particle to the force exerted on its surface.

We now assume an arbitrary time-dependent external force~$\vect{F}_2$ to be acting on the spherical particle positioned at~$\X_2$.
The zeroth-order solution for the translational velocity of the solid particle can readily be obtained from the Stokeslet solution as~$\vect{V}_2^{(0)} = \mu_0 \vect{F}_2$, where $\mu_0 = 1/(6\pi\eta b)$ is the usual Stokes mobility for a sphere moving in an unconfined viscous fluid.
The leading-order correction to the hydrodynamic self-mobility can be calculated from the reflected flow field as
\begin{equation}
	\left. \vect{v}^* \right|_{\X=\X_2} = \Delta \mu \vect{F}_2 \, . 
	\label{deltaMuDef}
\end{equation}
The latter result is often denominated as the mobility correction in the point-particle approximation~\cite{bickel06,bickel07}.
Higher-order correction terms can be obtained by employing a combination of the multipole expansion and the \Faxen~theorem~\cite{swan07,swan10}.
However, we will show in the sequel that this approximation, despite its simplicity, can surprisingly lead to a good prediction of the mobility correction when comparing with fully-resolved computer simulations.

By making use of the relations
\begin{subequations}
	\begin{align}
		\left. \boldNabla \psi_{n-1} \right|_{\R=\vect{x}_2} &= -\frac{n(n+1)}{2R^{n+2}} \, \vect{F}_2 \, , \\
		\left. \bgamma_{n-1} \right|_{\R=\vect{x}_2} &= -\frac{n}{R^{n+1}} \, \vect{F}_2 \, , \\
		\left. \left( \vect{t} \times \R \right) \varphi_{n-1} \right|_{\R=\vect{x}_2} &= \frac{\vect{F}_2}{R^{n-1}} \, , \\ 
		\left. \vect{r} \psi_{n-1} \right|_{\R=\vect{x}_2} &= \vect{0} \, , 
	\end{align}
\end{subequations}
in addition to inserting eq.~\eqref{insideSolution_finalized} into eq.~\eqref{deltaMuDef}, we write the scaled mobility correction  as
\begin{equation}
	\begin{split}
			\frac{\Delta\mu}{\mu_0} = 
			\frac{3b}{4} \sum_{n=1}^\infty
			 \bigg( 
			&-\frac{(n+1)(n+3)}{4} \, R^{3} a_n \\
			&- \frac{n+1}{2} \, R b_n + (n-1)c_n	
			 \bigg) R^{n-2} \, , \label{mobilityEq}
	\end{split}
\end{equation}
where we have used the relation $P_n'(1)=n(n+1)/2$ for the derivative at the end point.
We further remark that $R \in [0,1)$ because all distances have been scaled by the cavity radius~$a$.
The general term in the latter series, which we denote by $f_n(\alpha, R)$, has an asymptotic behavior at infinity that does not depend on the shear and bending properties of the membrane.
Specifically, we obtain as~$n\to\infty$
\begin{equation}
	f_n(\alpha, R) = \frac{3b}{16} \, n^2 \left(1-R^2\right)^2 R^{2n-2}
	 + \mathcal{O} \left( nR^{2n} \right) \, .
\end{equation}

In particular, for~$R=0$, the mobility correction simplifies to
\begin{equation}
	\left. \frac{\Delta\mu}{\mu_0} \right|_{R=0} = -\frac{3b}{4} \left( b_1-c_2 \right) 
	= -\frac{5b}{4} \frac{\alpha (2\lambda-1)}{5 + \alpha (2\lambda-1)} \, ,
\end{equation}
in full agreement with the result obtained in part~I for a particle concentric with the elastic cavity. 
We recall that the shear number~$\alpha$ has previously been defined by eq.~\eqref{alpha}, and the  dimensionless parameter~$\lambda$ associated with the Skalak ratio has been defined by eq.~\eqref{lambda}.

In the quasi-steady limit of vanishing frequency, the scaled correction to the mobility reads
\begin{equation}
	\lim_{\alpha\to\infty} \frac{\Delta\mu}{\mu_0} =
	\frac{\Delta\mu_\mathrm{R}}{\mu_0} + b \left( 1+\frac{3R^2}{4} \right) \, , 
	\label{mobilityInfShear}
\end{equation}
wherein~$\Delta\mu^\mathrm{R}/\mu_0$ is the scaled correction to the particle mobility associated with asymmetric motion inside a rigid spherical cavity.
This correction can readily be obtained by substituting the series coefficients given by eq.~\eqref{anbncnRigid} into eq.~\eqref{mobilityEq} to obtain
\begin{equation}
	\frac{\Delta\mu_\mathrm{R}}{\mu_0} = \sum_{n=1}^{\infty}
	\lim_{\alpha\to\infty} \frac{\Delta\mu}{\mu_0}
	= - \frac{9b}{16}	\frac{4-3R^2+R^4}{1-R^2} \, ,
\end{equation}
in agreement with the results by Aponte-Rivera and Zia~\cite{aponte16, aponte18, aponte-thesis}, who provided the elements of the grand mobility tensor for general motion inside a rigid cavity.
Interestingly, the particle mobility in the limit of infinite stiffness is found to be always larger than that inside a rigid cavity with no-slip velocity boundary condition on its interior surface.
Mathematically, this behavior can be justified by the fact that the limit and sum operators cannot generally be swapped in every situation.
In fact, using Fatou's Lemma~\cite{carothers00}, it can be shown that
\begin{equation}
	\lim_{\alpha\to\infty} \sum_{n=1}^{\infty} \left| f_n(\alpha, R) \right| \ge 
	\sum_{n=1}^{\infty} \lim_{\alpha\to\infty} \left| f_n(\alpha, R) \right| \, .
\end{equation}
That is, evaluating the sum over~$n$ before taking the limit~$\alpha\to\infty$ (as for an elastic cavity) could lead, under some circumstances, to a larger value in magnitude compared to the case in which the sum is taken after taking the limit (as it is the case for a rigid cavity).
This is explained by the fact that the dominated convergence theorem does not apply for the series function at hand~\cite{billingsley13}.

We further mention that the same limit given by eq.~\eqref{mobilityInfShear} is obtained when the cavity membrane only possesses resistance toward shear.
In the limit of infinite cavity radius, the classic result for motion parallel to a planar hard wall is recovered.
Specifically,
\begin{equation}
	\lim_{a\to\infty} \frac{\Delta\mu_\mathrm{R}}{\mu_0} =
	-\frac{9}{16} \frac{b}{h} \, , 
\end{equation}
wherein~$h=1-R$ denotes the distance between the center of the particle and the closest point of the cavity membrane.

Next, we consider an idealized cavity membrane with pure bending resistance and calculate the correction to the self-mobility function in the limit of~$\alphaB\to\infty$, corresponding to an infinite bending modulus or to particle motion in the quasi-steady limit of vanishing frequency.
After some algebra, we obtain
\begin{equation}
	\lim_{\alphaB\to\infty} \frac{\Delta\mu}{\mu_0} =
	\frac{\Delta\mu_\mathrm{D}}{\mu_0} + \frac{3b}{40} \left(5-2R^2\right)^2 \, , 
	\label{mobilityInfBending}
\end{equation}
wherein~$\Delta\mu_\mathrm{D}/\mu_0$ is the scaled correction to the particle mobility for motion inside a spherical drop of infinite surface tension (with vanishing normal velocity on its surface), given by
\begin{equation}
	\frac{\Delta\mu_\mathrm{D}}{\mu_0} =
	\sum_{n=1}^{\infty} \lim_{\alphaB\to\infty} \frac{\Delta\mu}{\mu_0}
	=  - \frac{3b}{32} \frac{20-15R^2-3R^4}{1-R^2}  \, .
\end{equation}

Again, the particle mobility in the vanishing-frequency limit for a membrane with pure bending is found to be always larger than that inside a spherical drop.
Notably, the mobility correction vanishes in the concentric configuration corresponding to~$R=0$ where the system behavior is solely determined by membrane shear properties.
This is in agreement with the results of part~I obtained by exactly solving the fluid motion equations for an extended particle of finite size concentric with an elastic cavity.

In the limit of infinite cavity radius, we recover the mobility correction near a planar fluid-fluid interface, 
\begin{equation}
	\lim_{a\to\infty} \frac{\Delta\mu_\mathrm{D}}{\mu_0} =
	-\frac{3}{32} \frac{b}{h} \, , 
\end{equation}
in agreement with the result by Lee and Leal~\cite{lee79}.

\begin{figure}
	\includegraphics[scale=1]{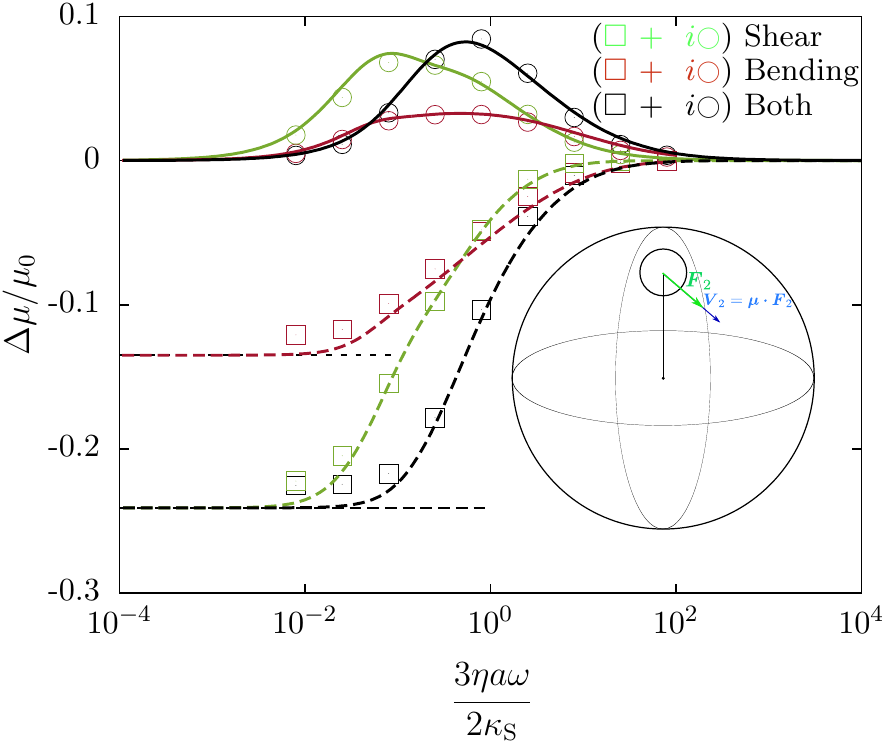}
	\caption{Variation of the correction to the self-mobility function inside a spherical elastic cavity (scaled by the bulk mobility) versus the scaled frequency. 
	The physical setup is sketched in the inset. 
	Squares ($\square$) and circles ($\fullmoon$) indicate the real and, respectively, imaginary parts of the mobility correction as obtained from the full boundary integral simulations performed for a cavity membrane endowed with pure shear (green), pure bending (red), or coupled shear and bending (black). 
	Solid and dashed lines give the corresponding analytical predictions (as described in the main text), which closely follow the numerical results. 
	Thin black horizontal dashed lines represent the vanishing-frequency limits.
	Here, $b=1/10$, $R=4/5$, and $\kB/(\kS a^2) = 2/75$.}
	\label{deltaMu_PARA}
\end{figure}

In the following, we assess the appropriateness and validity of our analytical calculations by direct comparison with computer simulations based on a completed-double-layer boundary integral method~\cite{pozrikidis01}.
The method is perfectly suited for solving numerically diverse flow problems in the Stokes regime involving both rigid and elastic boundaries.
For technical details regarding the computational method and its numerical implementation, we refer the reader to refs.~\onlinecite{daddi16b} and~\onlinecite{guckenberger16}.

To probe the effect of the confining elastic cavity on the motion of an encapsulated particle, we present in fig.~\ref{deltaMu_PARA} the variations of the scaled correction to the self-mobility as a function of the forcing frequency, for a cavity membrane possessing only shear (green), only bending (red), or both shear and bending deformation modes (black). 
Here, the particle of radius~$b=1/10$ is positioned at $R=4/5$ from the cavity center. 
We observe that the real (reactive) part of the mobility correction (shown as dashed lines) is a monotonically increasing function with frequency and approaches zero for larger forcing frequencies.
In contrast to that, the imaginary (dissipative) part (shown as solid lines) exhibits the typical bell-shaped profile which peaks at around $\omega \sim \kS/(\eta a)$.
In the low-frequency regime, the mobility correction approaches the plateau values predicted by eqs.~\eqref{mobilityInfShear} and \eqref{mobilityInfBending} for a cavity membrane with only shear elasticity or pure bending, respectively.
Overall, there is strong quantitative agreement between the full numerical solutions (symbols) and the theoretical predictions.
The small observed discrepancy notably for the real part in the low-frequency regime is most probably due to the finite size effect, because the analytical predictions are based on the point-particle approximation, whereas the numerical simulations necessarily account for the finite radius of the solid particle.

\section{Cavity motion and membrane deformation}
\label{sec:cavityMotionUndMemDeformation}

\subsection{Pair (composite) mobility}

\begin{figure}
	\includegraphics[scale=1]{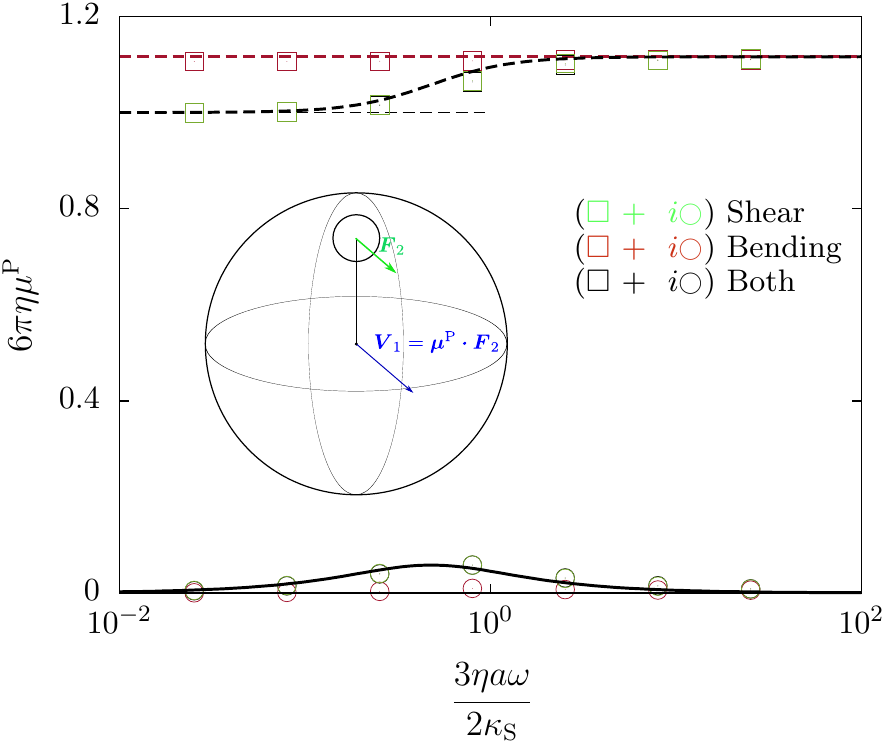}
	\caption{Variation of the scaled pair mobility function (bridging between the translational velocity of the cavity and the external force exerted on the solid particle, as sketched in the inlet) as a function of the scaled frequency.
	The theoretical predictions are shown as dashed and solid lines, for the reactive and dissipative parts, respectively.
	Symbols (same as fig.~\ref{deltaMu_PARA}) represent the boundary integral simulations results.
	The other system parameters are the same as in fig.~\ref{deltaMu_PARA}.
	In this plot, the pure-shear data points (green) mostly overlap with those for coupled shear and bending (black).}
	\label{deltaMuCaps_PARA}
\end{figure}

The hydrodynamic self-mobility discussed in sect.~\ref{sec:mobility} represents the particle response function to an external force.
In this regard, one can also define an analogous response function for the whole elastic cavity and its interior, that relates the translational velocity~$\vect{V}_1$ of the cavity centroid to the force~$\vect{F}_2$ exerted on the encapsulated particle via \mbox{$\vect{V}_1 = \vect{\mu}^\mathrm{P} \cdot \vect{F}_2$}.
In accordance to part~I, we call the tensor $\vect{\mu}^\mathrm{P}$ the pair (composite) mobility.
By symmetry, $\vect{V}_1 \parallel \vect{F}_2$~holds, so that the components of $\vect{\mu}^\mathrm{P}$ reduce to a single entry $\mu^\mathrm{P}$ connecting the corresponding magnitudes via~$V_1 = \mu^\mathrm{P} F_2$.

Without loss of generality, we assume in the following that~$\vect{F}_2$ is exerted along the $x$-direction.
Accordingly, the translational velocity of the elastic cavity can be calculated by integration over the fluid domain inside the cavity as~\cite{felderhof14}
\begin{equation}
	V_1(\omega) = \frac{1}{\Omega} \int_0^1 \!\!\!\! \Intd r \int_0^{2\pi} \!\!\!\!\!\!\Intd \phi \int_0^\pi \!\!\!\! \Intd \theta \,
	v_x^{(i)} (r,\phi,\theta,\omega) \, r^2 \sin\theta \, ,
\end{equation}
where~$\Omega = 4\pi/3$ is the scaled volume of the undeformed cavity, and 
\begin{equation}
	v_x^{(i)} = \left( v_r^{(i)} \sin\theta  + v_\theta^{(i)} \cos\theta \right) \cos\phi
	 - v_\phi^{(i)} \sin\phi \,  .
\end{equation}
The resulting frequency-dependent pair mobility function is obtained as
\begin{equation}
	\mu^\mathrm{P} = -\frac{1}{8\pi\eta} 
	\left( \frac{4R^2}{5} - 2 + a_1 + b_1 - c_2 \right) \, , 
\end{equation}
so that only the term corresponding to~$n=1$ remains after volume integration.
Upon simplification and rearrangement, the result can be presented in a scaled form as
\begin{equation}
	6\pi\eta \mu^{\mathrm{P}} = 
	\frac{3}{2} - \frac{3}{5} \, R^2 - \frac{5-6R^2}{10} \frac{\alpha \left(2\lambda-1\right)}{5+\alpha \left(2\lambda-1\right)} \, ,
\end{equation}
where the parameters~$\alpha$ and~$\lambda$ are defined by eq.~\eqref{alpha} and eq.~\eqref{lambda}, respectively.

Consequently, $\mu^\mathrm{P}$ depends only on the membrane shear properties and can be described by a simple Debye model with a single relaxation time $\tau / \tau_\mathrm{S} = 15 / \left( 2(2\lambda-1) \right)$, where~$\tau_\mathrm{S} = a\eta/\kS$ is a characteristic time scale for shear. 
Remarkably, the pair mobility can also become independent of frequency for~$R=\sqrt{30}/6 \approx 9/10$, a value for which~$6\pi\eta \mu^\mathrm{P} = 1$.
Nevertheless, as~$R \sim 1$, it becomes essential to ensure that the inequality~$R+b \ll 1$ remains satisfied, for the point-particle approximation employed here to be applicable.

In~fig.~\ref{deltaMuCaps_PARA}, we show the variations of the pair mobility (scaled by~$6\pi\eta$) as a function of the scaled frequency. 
Results are shown for a cavity membrane with pure shear (green), pure bending (red), and both shear and bending (black).
The pair mobility for a bending-only membrane remains unchanged upon varying the frequency and amounts to~$3/2-3R^2/5$.
In contrast to that, the reactive part for a membrane possessing a shear resistance shows a logistic sigmoid curve varying between 1 (when~$\alpha \to \infty$) and $3/2-3R^2/5$ (when~$\alpha = 0$), whereas the dissipative part exhibits a Gaussian-like -- or -- bell-shaped profile.
In all cases, there is strong agreement between the series-expansion theory (solid lines) and the full numerical solutions (symbols), confirming our theoretical predictions that the pair mobility is solely dependent on membrane shear properties (and independent of bending properties).

In analogy to the above discussion of the translational motion of the cavity in the presence of a force acting on the enclosed particle, 
one can also consider the corresponding rotational response.
The angular velocity~$\vect{\Omega}$ of the cavity is (due to symmetry) of the form~$\vect{\Omega}=\Omega \vect{e}_y$  
and has to fulfill $\vect{v}^{(q)}(\vect{r}) = \vect{\Omega} \times \vect{r}$
at the surface of the cavity~($r=1$), with $q \in \{ i, o \}$.
After some algebra, one obtains
\begin{equation}
 \Omega =\, \frac{3}{4}  \left< \vect{e}_y \cdot \left.\left( \vect{r} \times \vect{v}^{(q)} \right)
\right|_{r=1} \right> =\, \frac{F R}{8 \pi \eta}
\end{equation}
upon inserting our solution for the flow field, with angular brackets again denoting the surface average defined in eq.~(\ref{averagingDefinition}). 
Here, we find the same value of~$\Omega$ for the different series coefficients obtained for pure shear, pure bending, as well as combined shear and bending.
Additionally, we note that only the term~$\propto \boldsymbol{\Gamma}_0$ in eqs.~(\ref{fluid_flows}) contributes to the rotation of the cavity, while all other terms lead to vanishing contributions.
Accordingly, the angular velocity effectively stems only from the Stokeslet solution and does not depend on membrane shear and bending properties.

\begin{figure}[ht!]
	\includegraphics[scale=1]{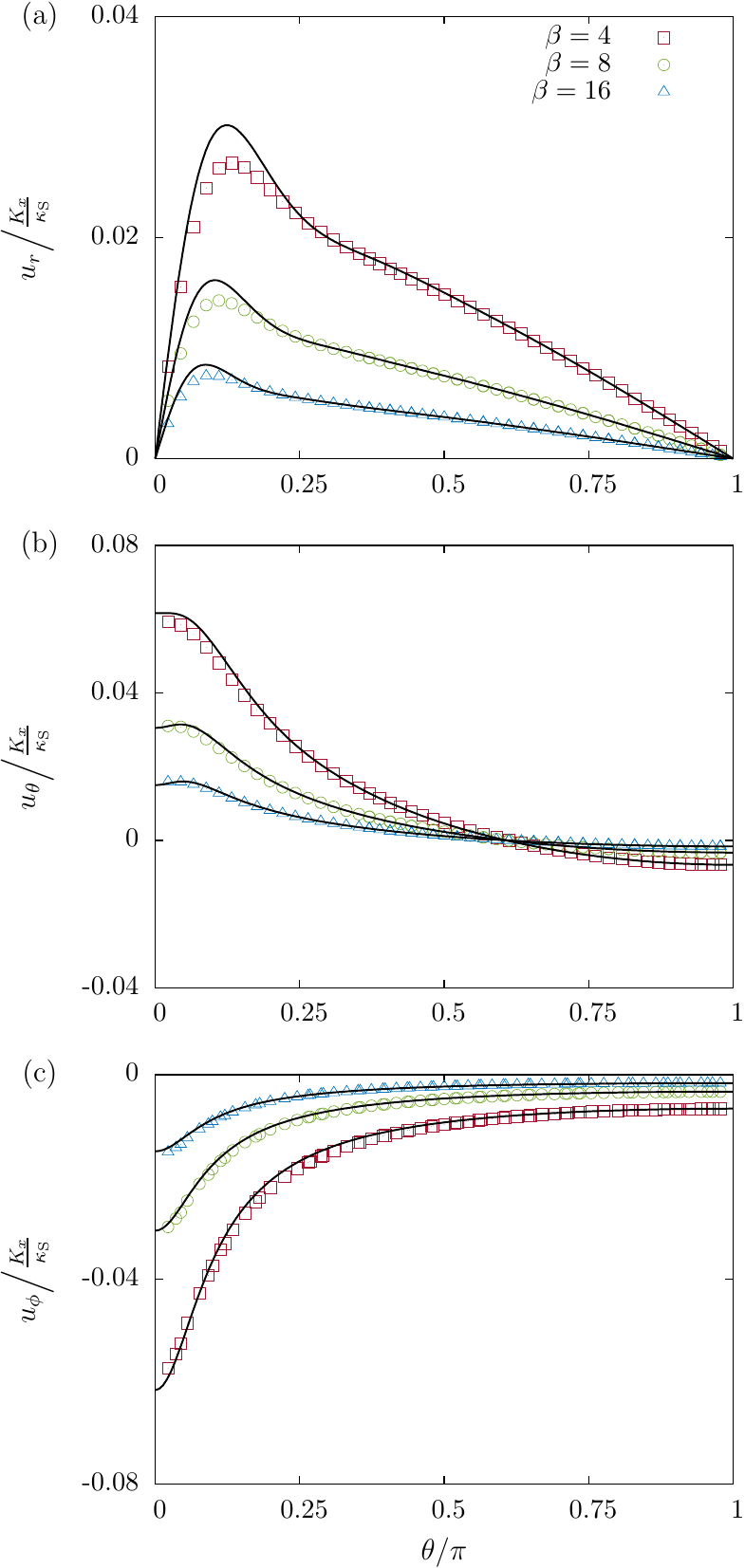}
	\caption{Scaled membrane displacement field. 
	$(a)$ Radial, $(b)$ circumferential, and~$(c)$ azimuthal components of the displacement field as a function of the polar angle $\theta$ for three scaled forcing frequencies [with $\beta=(3 \eta a \omega)/(2 \kS)$], evaluated at quarter oscillation period for~$t\omega_0=\pi/2$.
	The components of the local displacement fields are shown for their respective planes of maximum deformation ($\phi=0$ for $u_r$ and $u_\theta$ and $\phi=\pi/2$ for $u_\phi$). 
	Numerical results obtained for coupled shear and bending are shown as symbols (as indicated in the legend), while the solid lines represent the corresponding, closely matching theoretical predictions.}
	\label{memShape}
\end{figure}

\subsection{Membrane deformation}

The elastic deformation of the membrane can be assessed by calculating the displacement field $\vect{u} (\theta,\phi, \omega)$ resulting from the external force acting on the particle.
This field quantifies the motion of the material points of the cavity membrane relative to their initial positions in the undeformed state.
In the small deformation regime, the displacement field can readily be obtained from the no-slip boundary condition given by eq.~\eqref{no-slip_Frequency}, to obtain
\begin{align}
	8\pi\eta i\omega \, u_r &= \infSumOne 
				\left( -\frac{n}{2} \, A_n + B_n - C_{n+1} \right) \psi_{n-1} \, , \notag \\
	8\pi\eta i\omega \, \boldsymbol{\Pi} \vect{u} &= \infSumOne  
		   \left( \frac{n-2}{2} A_n - B_n \right) \frac{\boldsymbol{\Psi}_{n-1}}{n+1} 
		   + \infSum C_{n+1} \boldsymbol{\Gamma}_n \, . \notag 
\end{align}

We now define the reaction tensor~$\vect{R} $, a frequency-dependent tensorial quantity relating the membrane displacement field of the cavity to the asymmetric point force as~\cite{bickel07}
\begin{equation}
	\vect{u} (\phi,\theta,\omega) = \vect{R} (\phi,\theta,\omega) \cdot \vect{F} (\omega) \, .   
\end{equation}
By considering a harmonic oscillation of the form \mbox{$\vect{F} = \vect{K} e^{i\omega_0 t}$}, of amplitude~$\vect{K}$ and frequency~$\omega_0$, the membrane displacement in real space can readily be obtained from inverse Fourier transform as~\cite{bracewell99}
\begin{equation}
	\vect{u} (\phi,\theta, t) = \vect{R} (\phi,\theta,\omega_0) \cdot \vect{K} e^{i\omega_0t } \, .   
\end{equation}

An exemplary displacement field is displayed in fig.~\ref{memShape} as a function of the polar angle for three different forcing frequencies.
The azimuthal angle $\phi$ is chosen to represent the planes of maximum deformation for the respective components, as described in the figure caption.
Here, the cavity membrane is endowed with both shear and bending rigidities.
We observe that the radial component $u_r$ vanishes at the upper pole and shows a peak around $\theta/\pi \approx 1/8$, before decaying quasi linearly to zero upon increasing $\theta$.
The in-plane displacements~$u_\theta$ and~$u_\phi$ display a maximum value at the upper pole, and monotonically decay as~$\theta$ increases.
Our analytical predictions are in good agreement with numerical simulations.
Notably, we observe a small deviation in the plot for $u_r$ shown in panel~$(a)$ which is most probably due to a finite-sized effect. 
In contrast to the axisymmetric case discussed in part~I, the deformation here is (in general) largest in the tangential direction.

In typical biological situations, the forces that could be exerted by optical tweezers on particles are of the order of 1~pN~\cite{cipparrone10}.
The spherical cavity may have a radius of $10^{-6}~$m and a shear modulus of $\kS = 5 \times 10^{-6}~$N/m~\cite{Freund_2014}. 
For a scaled frequency $(3 \eta a \omega)/(2 \kS) = 4$, the membrane cavity is expected to undergo a maximal deformation of only about 1~\% of its initial undeformed radius.
Consequently, cavity deformations and deviations from the spherical shape are notably small.

\section{Conclusions}
\label{sec:conclusions} 

In summary, we have presented an analytical theory to describe the low-Reynolds-number motion of a spherical particle moving inside a spherical membrane cavity endowed with both shear elasticity and bending rigidity.
Here, we have focused on the situation in which the force exerted on the particle is directed tangent to the surface of the cavity.
Together with the axisymmetric results obtained in an earlier paper~\cite{daddi18cavity}, the solution of the elastohydrodynamic problem for a point force acting inside a spherical elastic cavity is thus obtained.

We have expressed the solution of the flow problem using the method of images.
For this purpose, the hydrodynamic flow field is represented by a multipole expansion, summing over modes in terms of spherical harmonics, in analogy with familiar methods in electrostatics.
The unknown series coefficients associated with each mode have been determined analytically from the prescribed boundary conditions of continuity of the fluid velocity field at the membrane cavity and discontinuity of hydrodynamic stresses as derived from Skalak and Helfrich elasticity models, associated with shear and bending deformation modes, respectively.

We have then explored the role of confinement on the motion of the encapsulated particle by calculating the frequency-dependent mobility functions.
The latter linearly couple the translational velocity of the particle to the external force exerted on it.
In the quasi-steady limit of vanishing actuation frequency, we have demonstrated that the hydrodynamic mobility inside a spherical elastic cavity is always larger than that predicted inside a rigid cavity of equal size with no-slip surface conditions.
In addition, we have quantified the translational and rotational motion of the confining cavity, finding that the translational pair (composite) mobility is uniquely determined by membrane shear elasticity and that bending does not play a role in the dynamics of the cavity.
We have further assessed the membrane deformation caused by the motion of the particle, showing that the cavity membrane primarily experiences deformation along the tangential direction. 

Finally, we have assessed the appropriateness and applicability of our theoretical approach by supplementing our analytical calculations with fully-resolved computer simulations of truly-extended particles using the boundary integral method.
Good agreement is obtained between theoretical predictions and numerical simulations over the full range of applied forcing frequencies. 
The developed method may find applications in the simulation of hydrodynamically interacting microparticles confined by a spherical elastic cavity,
or medical capsules that are directed to a requested site by magnetic forces acting on incorporated magnetic particles.

\begin{acknowledgments}
We would like to thank Stephan Gekle and the Biofluid Simulation and Modeling group at the University of Bayreuth, where the boundary integral code used in this work has been developed.
The authors gratefully acknowledge support from the DFG (Deutsche Forschungsgemeinschaft) through the projects LO~418/17-2 (H.L.), ME~3571/2-2 (A.M.M.), and DA~2107/1-1 (A.D.M.I).
\end{acknowledgments}

\section*{Author contribution statement}

A.D.M.I. conceived the study and performed the numerical simulations. 
C.H. and A.D.M.I. carried out the analytical calculations and drafted the manuscript.
C.H., H.L., A.M.M., and A.D.M.I. discussed and interpreted the results, edited the text, and finalized the manuscript.

% \newpage

\appendix*

\begin{widetext}
\section{Expression of the coefficients}

In this Appendix, we provide explicit expressions for the coefficients stated in eqs.~\eqref{AnBnCnShear} and~\eqref{AnBnCnBending} of the main body of the paper.
For an idealized membrane with pure shear, the coefficients are given by
\begin{align}
 K_1 &=\alpha\lambda \, n \, ( n-1 )\,  ( n+3 ) \, ( n+2 )\, , \notag  \\
 K_2 &=  ( n+1 ) \, ( 2n+1 ) \, \big( { 
\alpha }{n}^{4}\lambda+4\alpha {n}^{3}\lambda % \notag\\
+ ( 8-2
 \alpha +5\alpha \lambda ) \, {n}^{2}+ ( 16-4\alpha +2 \alpha \lambda ) \, n + 6 \big) \, , \notag \\
K_3 &= n \, \big( 4\,\alpha \lambda {n}^{5}  + ( -{\alpha }^{2}+10
\alpha \lambda+16+2\,{\alpha }^{2}\lambda ) \, {n}^{4} % \notag\\
+
 ( 32+6\alpha \lambda+4\alpha^{2}\lambda-2
 \alpha^{2} ) \, {n}^{3} % \notag\\
 + ( {\alpha }^{2}-2
\alpha^{2}\lambda+8-\alpha \lambda ) \, {n}^{2} \notag\\
&\quad- ( 4\alpha^{2}\lambda+\alpha \lambda-2{\alpha }^{2}+8+6
\alpha  )\, n-3-3\alpha  \big) \, , \notag \\
 K_4 &= n \, \left( \alpha {n}^{3}\lambda+ ( \alpha \lambda+4-{
 \alpha } ) \, {n}^{2}-2\alpha n+2\alpha -1 \right) \, , \notag \\
K_5 &= 2 \left( \alpha \lambda {n}^{3}+ ( 3{ \alpha 
}\lambda+4 -\alpha ) \, {n}^{2}
+ ( 2\alpha \lambda-2{ 
\alpha }+8 ) \, n +3 \right) \, , \notag \\
K_6 &= -( n+1 ) \, n \, \big(-2\alpha {n}^{3}+ ( -8-15
\alpha +8\alpha \lambda ) \, {n}^{2} % \notag\\
+ (-43\alpha+28\alpha \lambda-16 ) \, n
-42\alpha +24\alpha\lambda-6 \big) \, , \notag \\
K_7 &=  ( n+2 ) \, ( \alpha {n}^{2}+(3\alpha +4) \,n+6 ) \, K_5 \, , \notag \\
K_8 &=-( n+1 ) \, ( n+2 )  \, ( 2n+1 )  \,
 ( \alpha {n}^{2}+(3 \alpha +4)\, n+6 ) \, , \notag \\
 K_9&= \tfrac{1}{4} \left( \alpha {n}^{3}+(2{
 \alpha }+4)\,{n}^{2}+(6-\alpha )\,n-2\alpha +2 \right) \, , \notag
\end{align}
where we recall that~$\alpha=2\kS/(3\eta i\omega)$ is the shear number, and~$\lambda=C+1$ is the dimensionless parameter associated with the Skalak ratio. 
For an idealized membrane with pure bending resistance, the corresponding coefficients read
\begin{align}
 Q_1 &= \alpha_\mathrm{B} \, n \, ( n+3 ) \,  ( n+2 )^2 \,( n-1 )^2 \, , \notag \\
 Q_2 &= - \alpha_\mathrm{B} \, n^6-3 \alpha_\mathrm{B} \, n^5+ \alpha_\mathrm{B} \,n^4
+7 \alpha_\mathrm{B} \, n^3+8  \, n^2 % \notag\\
   + \left( 16-4 \alpha_\mathrm{B} \right) \, n + 6 \, , \notag \\
 Q_3 &=   n  \, \big( 2 \alpha_\mathrm{B} \,n^6+6 \alpha_\mathrm{B} \,n^5-2
 \alpha_\mathrm{B} \, n^4+ ( 8-14 \alpha_\mathrm{B} ) \, n^3 % \notag\\
 +12  \, n^2+ ( -2+8 \alpha_\mathrm{B} )  \, n -3\big) \, , \notag \\
Q_4 &=n \, \big(\alpha_\mathrm{B} \, n^7 + 10\alpha_\mathrm{B}  \, n^6 + 17\alpha_\mathrm{B}  \, n^5 + (4-20\alpha_\mathrm{B})  \,n^4 % \notag\\
 + (40-40\alpha_\mathrm{B}) \,n^3 + (16\alpha_\mathrm{B}+47) \, n^2 % \notag\\
 + (-10+16\alpha_\mathrm{B}) \, n -12\big) \, , \notag \\
Q_5 &=-2  \, n  \, (n+1) \, (n+2) \,  \big(\alpha_\mathrm{B} \, n^4+4\alpha_\mathrm{B} \, n^3-3\alpha_\mathrm{B} \, n^2 % \notag\\
-(2+10\alpha_\mathrm{B}) \, n-1+8\alpha_\mathrm{B}\big) \, , \notag\\
Q_6 &= 2 \, (n+1)  \, \big(\alpha_\mathrm{B} \, n^6+2\alpha_\mathrm{B} \, n^5-3\alpha_\mathrm{B} \, n^4 % \notag\\
-(4\alpha_\mathrm{B}+2) \, n^3+(-21+4\alpha_\mathrm{B}) \, n^2-34 \, n-12\big) \, , \notag \\
Q_7 &=2 \, (n+2)^2 \, \big(\alpha_\mathrm{B} \, n^5+2\alpha_\mathrm{B} \, n^4-3\alpha_\mathrm{B} \, n^3 % \notag \\
+4 \, n^2-4\alpha_\mathrm{B} \, n^2+(4\alpha_\mathrm{B} \, +8) \, n+3\big) \, , \notag
\end{align}
wherein~$\alphaB = \kB/(\eta i\omega)$ denotes the bending number.

\end{widetext}

\end{document}